\documentclass{PoS}

\usepackage{graphicx}
\usepackage{amssymb}
\usepackage{fontenc}
\usepackage{mathptmx}
\usepackage{epsf}
\usepackage{psfig}

\title{Studying magnetic fields in several parsec-scale AGN jets using Faraday Rotation}

\ShortTitle{Studying magnetic fields in several parsec-scale AGN jets using Faraday Rotation}

\author{\speaker{Andrea Reichstein}%
        \thanks{Funding for this research was provided by Science Foundation Ireland.}\\
       University College Cork, Ireland\\
       E-mail: \email{amn.reichstein@gmail.com}}

\author{Denise Gabuzda\\
        University College Cork, Ireland\\
        E-mail: \email{d.gabuzda@ucc.ie}}

\abstract{We present multi-frequency radio observations from the Very
            Long Baseline Array (VLBA) of selected AGN that seem to have a B-field
            structure with a central ``spine'' of B-field orthogonal to the jet and a
            longitudinal B-field near one or both edges of the jet. Two explanations for
            this structure have been discussed in the literature: shocks making the
            central orthogonal field combined with a jet-medium interaction causing the
            longitudinal ``sheath'', or both components produced by a helical jet magnetic
            field. One way to investigate this structure is to look for gradients in the
            Faraday Rotation across the jet. We will discuss results for 0333+321, 1150+812 and 2037+511
            providing evidence for the latter picture.}

\FullConference{10th European VLBI Network Symposium and EVN Users Meeting: VLBI and the new generation of radio arrays \\
                September 20-24, 2010\\
                Manchester, UK}

\begin{document}

\section{Introduction}   
\begin{figure}[hbt]
\center
\includegraphics[width=5cm]{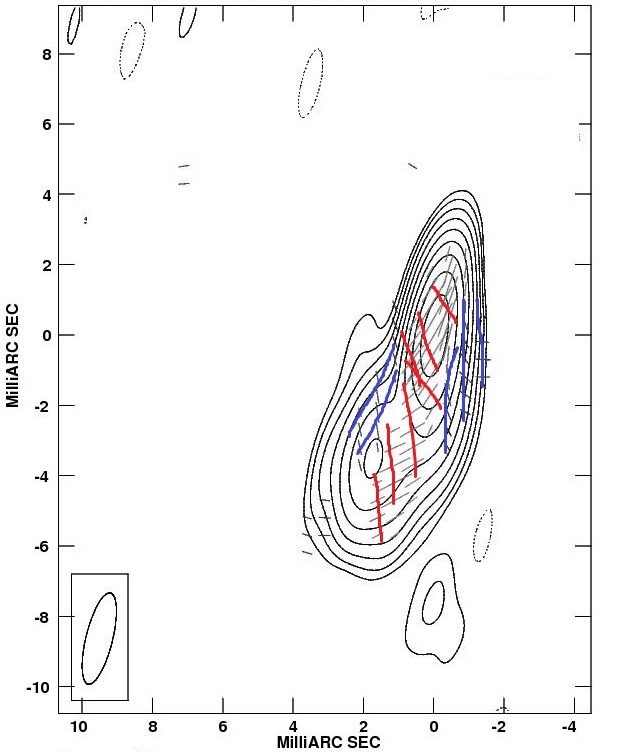}
\includegraphics[width=9cm]{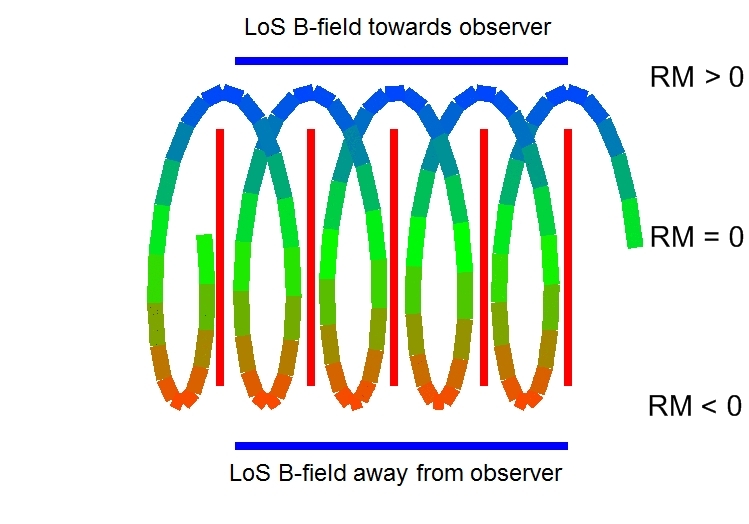}
\caption{Left: map of 1504-166, with counturs of the total intensity at 12.9 GHz, the sticks indicate the polarization angles. Assuming that the magnetic field is perpendicular to the polarization angle (optically thin jet) the red and blue lines indicate the orientation of the line-of-sight magnetic field, forming the so called ``spine-sheath'' structure. Right: Schematic illustration how a helical magnetic field (its projection on the sky) can appear as a ``spine-sheath' structure with the B field orthogonal to the jet axis in the center and longitudinal at the jet edges. The colour coding of the helix shows the expected behaviour of the observed Faraday Rotation gradient transverse to the jet. The viewing angle is near 90$^o$ to the jet axis in the jets' rest frame. Due to the different directions of the line-of-sight B field on either side of the jet, the rotation measure has opposite signs.}
\label{helix}
\end{figure}

When first discovered in the active galactic nucleus (AGN) 1055+018 \cite{1999ApJ...518L..87A}, ``spine-sheath'' polarization structure was quite unexpected. Early VLBI polarimetry experiments at 5GHz suggested that, depending on the jet's angle to the line-of-sight, the image might be dominated by the ``spine'' or by the ``sheath'', but wouldn't show both at once (see \cite{1999ApJ...518L..87A}). 
One possible explanation for an AGN with ``spine-sheath'' polarization structure is interaction of the jet with the surrounding medium, causing a deceleration of the plasma near the jet edges and stretching out the magnetic field along the jet, forming a shear layer. The polarization is strongest where the interaction is strongest. In this picture the orthogonal magnetic field is caused by a series of transverse shocks which are more highly polarized than regions between them \cite{1994ApJ...437..122W,1999ApJ...518L..87A}.\newline
\indent A  second  possible  explanation is  an overall  helical  magnetic  field along the jet, where the projected toroidal component of the field is dominant at the jet axis, and the longitudinal field component becomes dominant at the  jet edges \cite{2005MNRAS.360..869L,2005MNRAS.356..859P}. This explanation is simpler and does not require two independent mechanisms to work.
Figure \ref{helix} shows an example of ``spine-sheath'' polarization structure and how to explain such a structure with a helical magnetic field. \newline
\indent To investigate this phenomenon we are making use of the effect of Faraday Rotation.

\section{Faraday Rotation}
When a linearly polarized electromagnetic wave travels through magnetized plasma, the polarization angle $\chi$ rotates due to the different propagating speeds of left and right-circular polarized components of the wave. This is called Faraday Rotation. The amount of rotation is proportional to the square of the observing wavelength, ${\lambda}^2$, as well as the integral of the electron density and  the line-of-sight component of the B-field, while its sign is determined by the direction of the line-of-sight B-field:
	\[\chi = \chi_{0} + RM{\lambda}^2
\]
	\[RM \propto \int{n_{e} B \cdot dl}
\]
Where $\chi_{0}$ is the unrotated polarization angle, $n_{e}$ is the electron density and $B \cdot dl$ is the line-of-sight magnetic field.
We call the coefficient of ${\lambda}^2$ the Rotation Measure (RM). Having simultaneous multifrequency observations, the RM can be easily determined by measuring the polarization angle for each pixel at each frequency and performing a linear fit on these. This also gives us the intrinsic (unrotated) polarization angle.\newline
\indent Thus, systematic gradients in Faraday Rotation across the jets of AGN can be interpreted as representing a systematic change in the line-of-sight component of a helical/toroidal B-Field (Figure \ref{helix}).
\nopagebreak[4]
\section{Observations and results}
\begin{figure}[htb]
\center
\includegraphics[width=6.5cm]{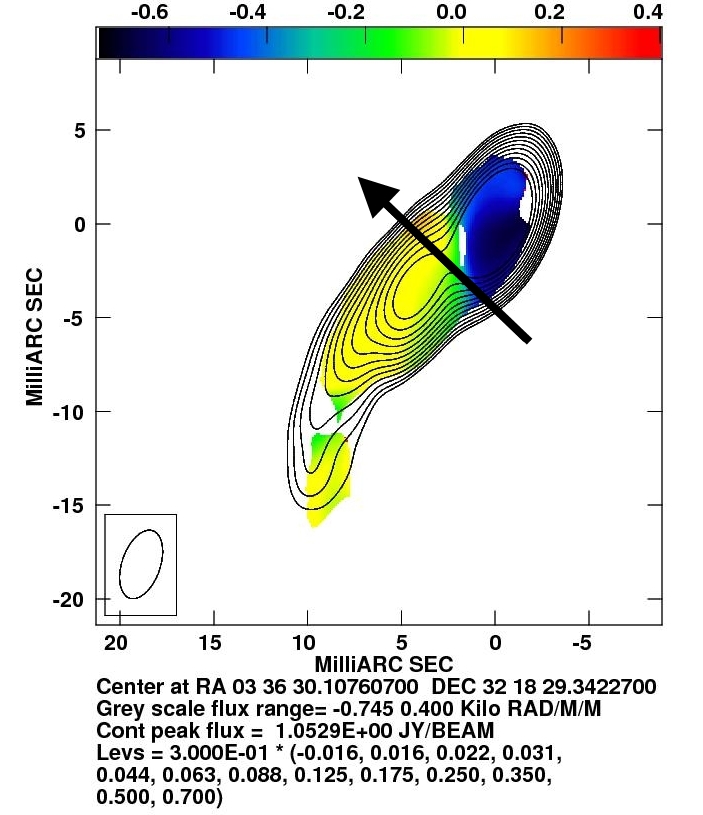}
\includegraphics[width=8.5cm]{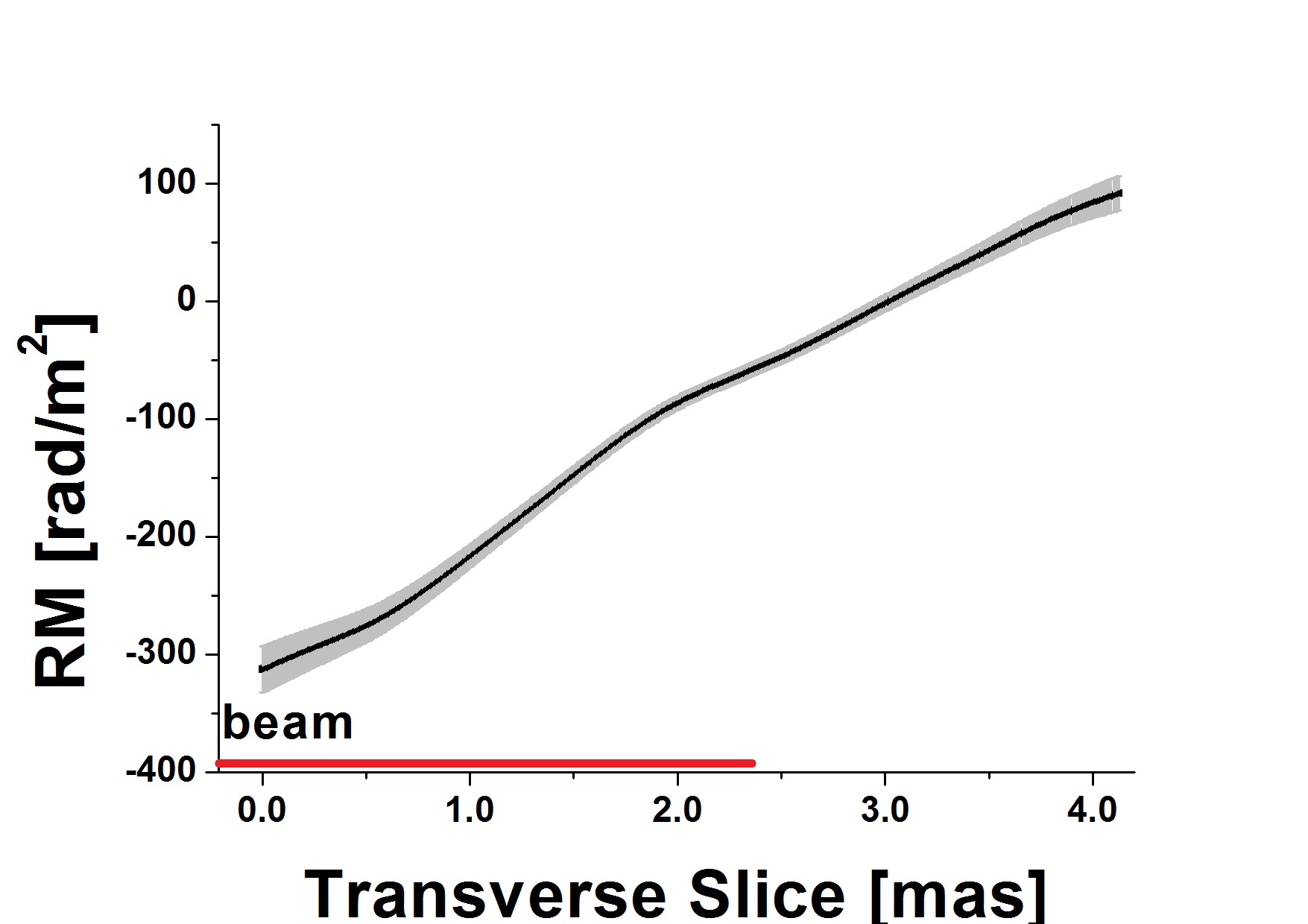}

\caption{RM map for 0333+321 from 4.6 to 8.9 GHz with contours of the 4.6 GHz total intensity map (left). The black line in the map indicates the slice across the RM map, which is shown in the right plot: the black line in the plot shows the RM and the gray area the error of the fit at each pixel. The beamsize is shown in the bottom of the plot (red line). The gradient is monotonic across the jet.}
\label{0333}
\end {figure}

\begin{figure}[htb]
\center
\includegraphics[width=7cm]{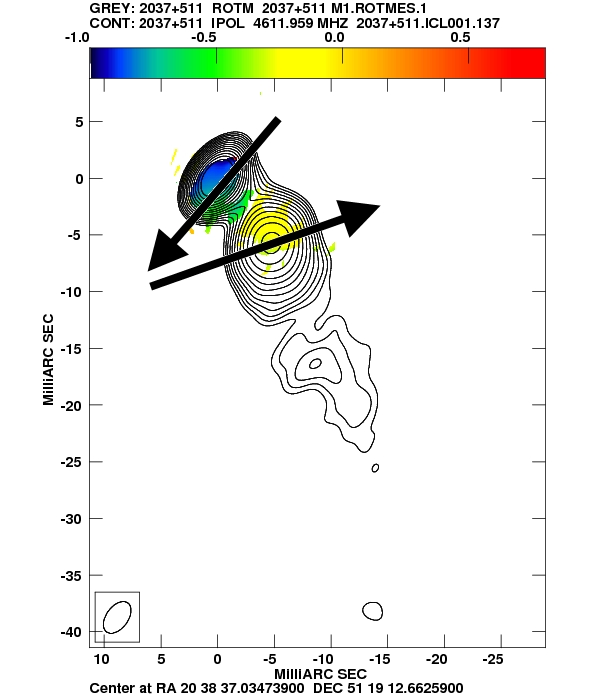}
\includegraphics[width=8cm]{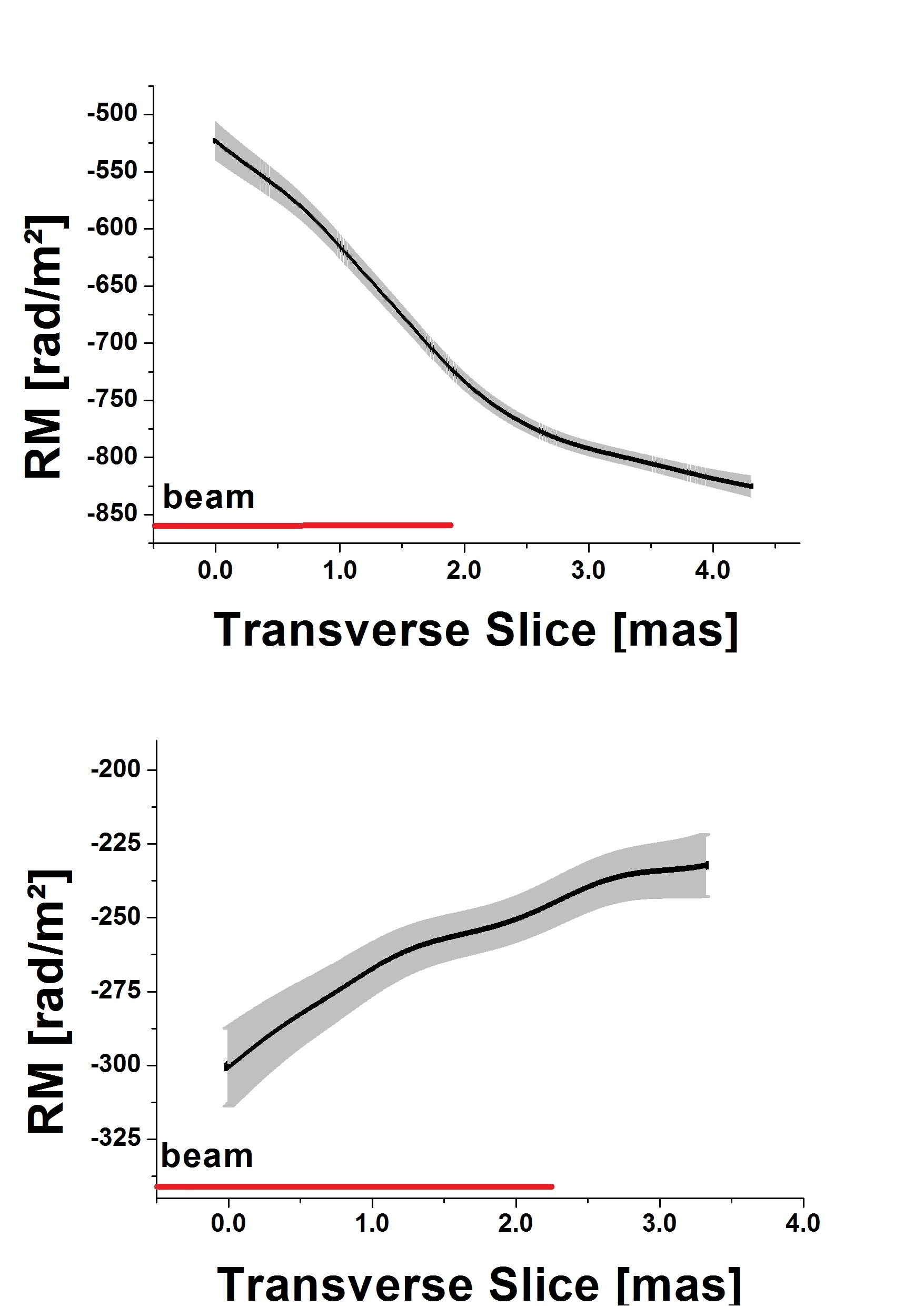}
\label{2037}
\caption{RM map for 2037+511 from 4.6 to 15.4 GHz with contours of the 4.6 GHz total intensity map (left). The black arrows in the map indicate the directions of the gradients, as well as the slices across the RM map which are shown in the right plots: the black line in the plots shows the RM and the gray area the error of the fit at each pixel. The beamsize is shown in the bottom of the plot (red line).The gradients are monotonic across the jet, but the direction switches along the jet.}
\end{figure}

24 AGN showing evidence for ``spine-sheath'' polarization structures were selected for multi frequency radio observations (4.6, 5.1, 7.9, 8.9, 12.9, 15.4 GHz) with the Very Long Baseline Array (VLBA). Here we present selected results for one of these two experiments. The observations were obtained with the Very Long Baseline Array (VLBA) on the 27th of September 2007. The data were calibrated and imaged in the NRAO AIPS package using standard techniques.\newline
\indent Analyzing their Rotation Measure (RM) distributions we have found some evidence for helical magnetic field structures in seven out of twelve objects. As an example, Figure \ref{0333} shows a transverse RM gradient across the entire resolved jet region of 0333+321. This is consistant with a helical/toroidal magnetic field structure surrounding the jet and confirms the results of Asada et al. \cite{2008ApJ...682..798A}, who found a very similar gradient in that source. The gradient of 0333+321 is slanted; this could be the result of a transverse RM gradient across and a decreasing RM along the jet, for example, due to a helical magnetic field combined with a fall-off of electron density/magnetic field strength with distance from the core.
\begin{figure}[htb]
\center
\includegraphics[width=7cm]{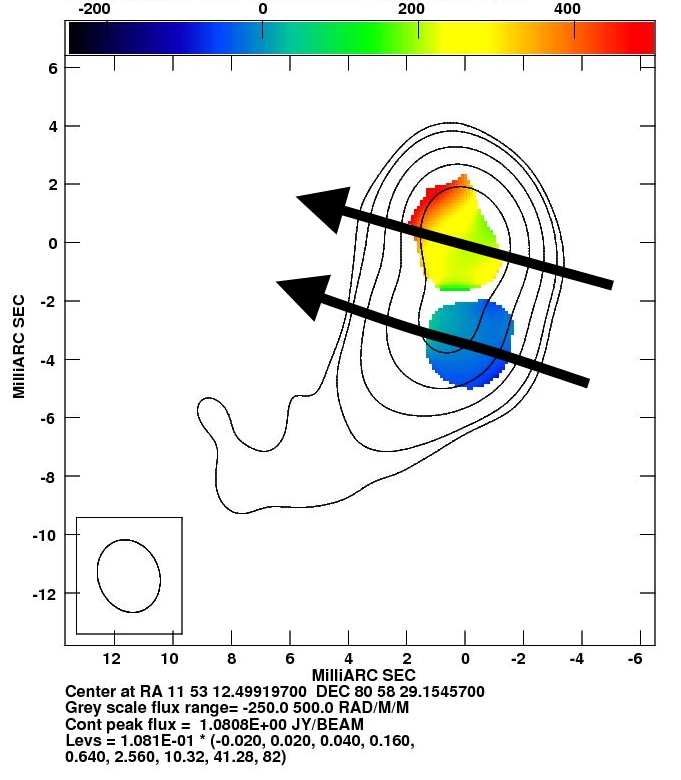}
\includegraphics[width=8cm]{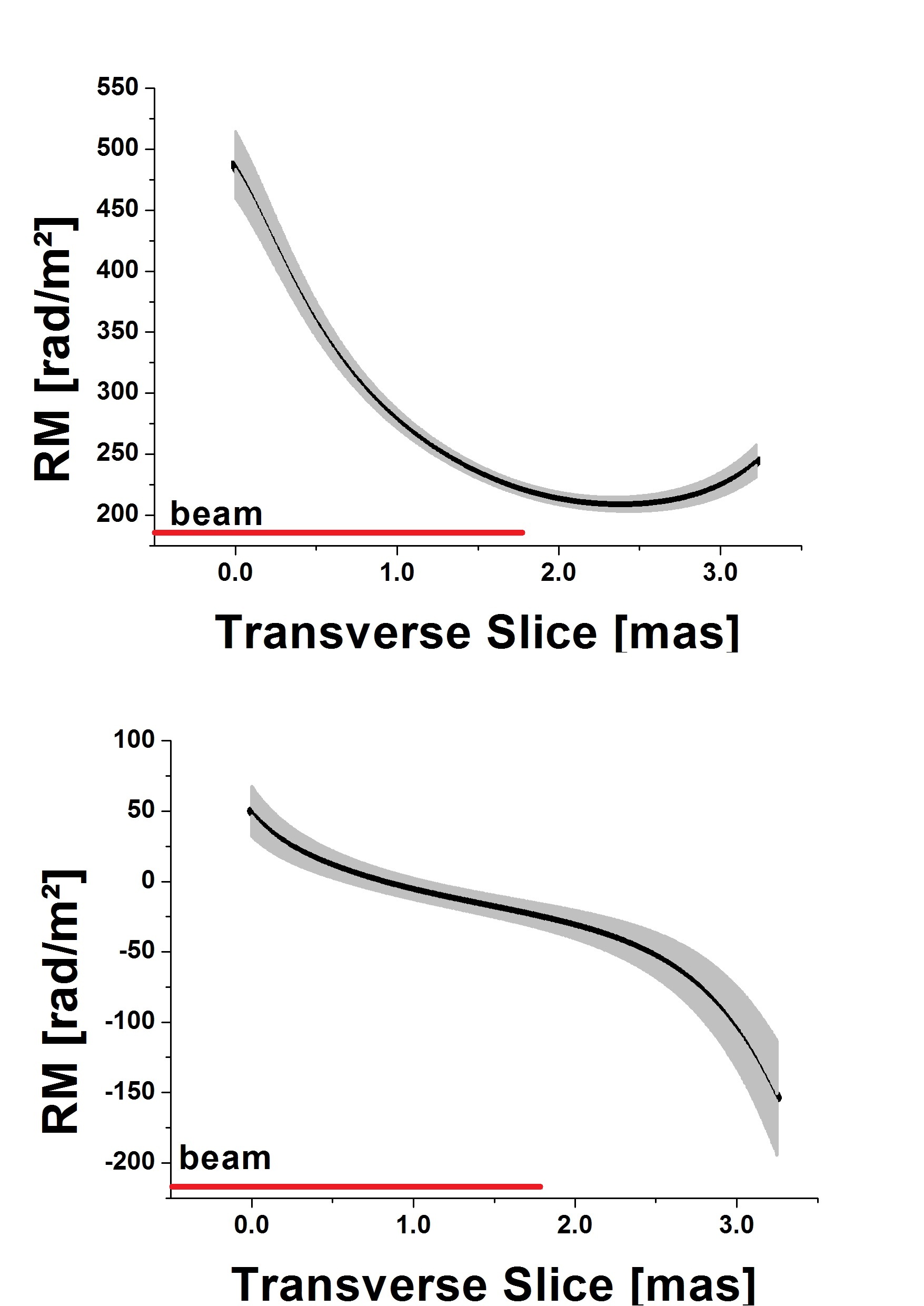}

\caption{RM map for 1150+812 from 4.6 to 8.9 GHz with contours of the 4.6 GHz total intensity map (left). The black arrows in the map indicate the slices across the RM map, which are shown in the right plot: the black line in each plot shows the RM and the gray area the error of the fit at each pixel. The beamsize is shown in the bottom of the plot (red line). Both plots show a gradient transverse to the jet.}
\label{1150}
\end {figure}
Another transverse RM gradient was found across the jet of 1150+812, see Figure \ref{1150}.\newline
\indent We have found three sources that show two oppositely directed transverse RM gradients at different distances from the core. They could be due to a ``nested-helix'' B-field structure, where the B-field emerging with the jet closes in the outer accretion disk \cite{2009arXiv0905.2368M}. In this picture the net RM gradient has contributions from both regions of helical field, and the dominant region determines the direction of the gradient. Figure \ref{2037} shows two oppositely directed transverse RM gradients in 2037+511, one in the core region and one about 10 mas from the core.\newline
\indent We have also found for our sources that the degree of polarization tends to be higher at the edges of the jet, which is expected for a helical magnetic field.

\nopagebreak[4]
\section{Discussion and Conclusion}
We have found tranverse rotation measure gradients across several new sources and were also able to confirm the gradient found by Asada et al. \cite{2008ApJ...682..798A} in 0333+321. We find some evidence for transverse RM gradients that may be associated with helical jet magnetic fields in 7 out of 12 ``spine-sheath'' AGN analysed. The simplest explanation for these transverse gradients is a helical magnetic field wraped around the jet, whose changing line-of-sight magnetic field components cause the observed gradients.
All our ``spine-sheath'' sources show an increase of fractional polarization towards the jet edges, which can also be explained by a helical magnetic field.
This provides evidence that their polarization structures are inherently assosiated with helical B-fields.\newline
\indent Our future work will include analysing 18-22cm VLBA polarization observations of several ``spine-sheath'' sources. We are also performing a Circular Polarization analysis using these data in collaboration with Vasiliy Vitrishchak, in order to search for connections between the circular polarization and RM gradients (see, e.g. \cite{2008MNRAS.384.1003G}).
\nopagebreak[4]
\acknowledgments Funding for this research was provided by Science Foundation Ireland.

\end{document}